\documentclass[prb,aps,floats,twocolumn]{revtex4}
\usepackage{epsfig,amssymb,amsmath}
\begin{document}
%-------------------------------------------------------------------
\title{Integrability and action operators in quantum Hamiltonian systems}  
\author{Vyacheslav V. Stepanov and Gerhard M{\"u}ller} 
\affiliation{Department of Physics, University of Rhode Island, Kingston RI
  02881-0817} 
\date{\today~--~1.6}
\begin{abstract}
  For a (classically) integrable quantum mechanical system with two degrees of
  freedom, the functional dependence $\hat{H}=H_Q(\hat{J}_1,\hat{J}_2)$ of the
  Hamiltonian operator on the action operators is analyzed and compared with the
  corresponding functional relationship $H(p_1,q_1;p_2,q_2) = H_C(J_1,J_2)$ in
  the classical limit of that system. The former is shown to converge toward the
  latter in some asymptotic regime associated with the classical limit, but the
  convergence is, in general, non-uniform.  The existence of the function
  $\hat{H}=H_Q(\hat{J}_1,\hat{J}_2)$ in the integrable regime of a parametric
  quantum system explains empirical results for the dimensionality of manifolds
  in parameter space on which at least two levels are degenerate. The
  comparative analysis is carried out for an integrable one-parameter two-spin
  model. Additional results presented for the (integrable) circular billiard
  model illuminate the same conclusions from a different angle.

\end{abstract}
\pacs{??}
\maketitle
%%%%%%%%%%%%%%%%%%%%%%%%%%%%%%%%%%%%%%%%%%%%%%%%
%
\section{Introduction}\label{sec1}
%
%%%%%%%%%%%%%%%%%%%%%%%%%%%%%%%%%%%%%%%%%%%%%%%%
A conspicuous phenomenological discriminant between quantized integrable and
nonintegrable parametric Hamiltonian systems with two or more degrees of freedom
is the occurrence or prohibition of level crossings between states within the
same invariant Hilbert subspace of the underlying symmetry
group.\cite{Gutz90,Reic92,Gutz98} Consider a quantum system with $d$ continuous
parameters whose classical counterpart is integrable on a manifold of
dimensionality $d_I \leq d$ in parameter space. Empirical evidence suggests that
level crossings occur on $(d_I-1)$-dimensional manifolds which are embedded in
the integrability manifold. A recent study,\cite{SM98} which investigated this
issue systematically, showed for a two-spin model with $d=6$ and $d_I=5$ that
the level crossing manifolds are, in fact, four-dimensional and that they are
all confined to the five-dimensional integrability manifold. It showed,
moreover, that the (classical) integrability manifold can be reconstructed from
the (intrinsically quantum mechanical) level crossing manifolds.

The focus of the present study is to illuminate the natural cause underlying
this characteristic relationship between level crossing manifolds and
integrability manifolds. We attribute this relationship to the presence of
action operators as constituent elements of the Hamiltonian operator for
integrable quantum systems.

The textbook solution of an integrable classical dynamical
system with two degrees of freedom, specified by an analytic function
$H(p_{1},q_{1};p_{2},q_{2})$ of canonical coordinates, is to transform the
Hamiltonian into a function of two action coordinates: $H=H_C(J_1,J_2)$. The
canonical transformation $(p_i,q_i)\to(J_i,\theta_i), i=1,2$ to action-angle
coordinates amounts to a solution of the dynamical problem because it transforms
Hamilton's equations of motion, $\dot{p}_i = -\partial H/\partial q_i$,
$\dot{q}_i = \partial H/\partial p_i$, generically a set of coupled nonlinear
differential equations, into $\dot{J}_i = 0$, $ \dot{\theta}_i = \partial
H_C/\partial J_i \equiv \omega_i$ with the solutions $J_i = {\rm const}$,
$\theta_i(t) = \omega_it + \theta_i^{(0)}$.

This solution is guaranteed whenever a second integral of the motion can be
found, i.e. an analytic function $I(p_{1},q_{1};p_{2},q_{2})$ which is
functionally independent of $H$ and has a vanishing Poisson bracket with $H$:
$dI/dt=\{H,I\}=0$. Deriving the expressions $H_C(J_1,J_2)$ and $I_C(J_1,J_2)$ from
$H$ and $I$ requires the use of separable canonical coordinates. Finding
separable coordinates can be a difficult task even if the second invariant is
known.

The functions $H_C(J_1,J_2)$ and $I_C(J_1,J_2)$ establish a pivotal link between
an integrable classical system and a quantized version of it.  Semiclassical
quantization derives its raison d'{\^e}tre from the obvious fact that quantizing a
functional relation is much less problematic if it involves only quantities such
as $H,I,J_1,J_2$ whose quantum counterparts are guaranteed to be commuting
operators.

%%%%%%%%%%%%%%%%%%%%%%%%%%%%%%%%%%%%%%%%%%%%%%%%
%
\section{Quantum versus quantized}\label{sec2}
%
%%%%%%%%%%%%%%%%%%%%%%%%%%%%%%%%%%%%%%%%%%%%%%%%%
In the context of this study, it is useful to distinguish and compare three
versions of the same model system:
(i) the {\it quantum} version, (ii) the {\it classical} version, and (iii) the
(semiclassically) {\it quantized} version.

The primary version is the quantum model, specified by the Hamiltonian expressed
as an operator valued function of a set of dynamical variables (position,
momentum, spin, $\ldots$) The commutation relations of these operators and the
metric of the associated Hilbert space along with the rules of quantum mechanics
then determine, via the Heisenberg equation of motion, the time evolution of any
observable quantity of interest.

The classical limit converts the Hamiltonian operator into the classical energy
function, the commutator algebra of dynamical variables into the symplectic
structure (the fundamental Poisson brackets), and the Heisenberg equation of
motion for any operator into the Hamilton equation of motion for the
corresponding classical quantity. These quantities, in turn, enable us to express
the energy function as a classical Hamiltonian, i.e.  as a function of canonical
coordinates.
 
The quantization of a classical Hamiltonian system requires a prescription for
translating the functional relations between classical dynamical variables into
functional relations between corresponding operators. Semiclassical quantization
is one neat and clean procedure applicable to all integrable classical systems.
It borrows from classical mechanics the functional dependence,
$\hat{H}=H_C(\hat{J}_1,\hat{J}_2)$, of the Hamiltonian on the action operators
and postulates that the eigenvalue spectrum of the latter consists of
equidistant levels spaced by $\hbar$:
\begin{equation}\label{sclac}
\langle\hat{J}_i\rangle = \hbar\left(n_i + \frac{1}{4}\alpha_i\right),~ ~ i=1,2
\end{equation}
with integer $n_i$. The (integer) Maslov indices $\alpha_i$ are determined by the
topology of the classical trajectories in phase space.\cite{Perc77}
Semiclassical quantization thus makes specific predictions for the energy level
spectrum of the quantized version of the model system at hand.

In general, the (semiclassically) quantized and the (primary) quantum energy
level spectra of one and the same integrable model system do not coincide. The
relationship between the two spectra will be investigated in Sec.~\ref{sec3} for
an integrable two-spin model and in Sec.~\ref{sec4} for the (integrable)
circular billiard model.

%%%%%%%%%%%%%%%%%%%%%%%%%%%%%%%%%%%%%%%%%%%%%%%
%
\section{Two-spin model}\label{sec3}
%
%%%%%%%%%%%%%%%%%%%%%%%%%%%%%%%%%%%%%%%%%%%%%%%
We consider two quantum spins ${\bf \hat{S}}_1, {\bf \hat{S}}_2$ of equal length
$\sqrt{\sigma(\sigma+1)}$ $(\sigma = \frac{1}{2}, 1, \frac{3}{2},\ldots)$
interacting via a uniaxially symmetric exchange interaction:\cite{note1}
\begin{equation}\label{Hxxz}
\hat{H} = -\left(\hat{S}_1^x\hat{S}_2^x + \hat{S}_1^y\hat{S}_2^y\right) 
-\kappa\hat{S}_1^z\hat{S}_2^z.
\end{equation}
The second integral of the motion, which follows from Noether's theorem, is
\begin{equation}\label{Mz}
\hat{I} = \hat{M}_z = \frac{1}{2}\left(\hat{S}_1^z + \hat{S}_2^z\right).
\end{equation}
In the classical limit $\hbar \to 0$, $\sigma \to \infty$,
$\hbar\sqrt{\sigma(\sigma+1)} = s$, the operators ${\bf \hat{S}}_i$ turn into
3-component vectors, ${\bf S}_i = s(\sin\vartheta_i \cos\varphi_i$,
$\sin\vartheta_i \sin\varphi_i$, $\cos\vartheta_i )$, and Eq.~(\ref{Hxxz}) then
describes the energy function of an autonomous Hamiltonian system with two
degrees of freedom and canonical coordinates $p_i= s
\cos\vartheta_i,~~q_i=\varphi_i,~~i=1,2$.\cite{MTWKM87}

%%%%%%%%%%%%%%%%%%%%%%%%%%%%%%%%%%%%%%%%%%%%%%%
%
\subsection{Classical actions}\label{sec3A}
%
%%%%%%%%%%%%%%%%%%%%%%%%%%%%%%%%%%%%%%%%%%%%%%%
Generically, the classical time evolution of this system is nonlinear and
quasiperiodic. In the parameter range $0<\kappa<1$, the following relation
between the integrals of the motion $H=E$ (energy), $I=M_z$ (magnetization) and
a set of classical actions $J_1,J_2$ can be inferred from the exact
solution:\cite{SM90}
\begin{equation}\label{xxzcla}
J_1 = 2M_z,~~ J_2 = \frac{1}{2\pi}\int_0^\tau dt
\frac{z\dot{\zeta}}{1+\zeta^2},
\end{equation}
\vspace*{-0.3cm}
\begin{eqnarray*}
z(t) &\equiv& \frac{1}{2}s(\cos\vartheta_1 - \cos\vartheta_2) 
= z_0{\rm sn}(\rho t,z_0/a), \\
\zeta(t) &\equiv& \tan(\varphi_1 - \varphi_2)
= \frac{\rho z_0{\rm cn}(\rho t,z_0/a){\rm dn}(\rho t,z_0/a)}
{E+\kappa[M_z^2 - z_0^2{\rm sn}^2(\rho t,z_0/a)]}, \\
z_0^2 &=& z_m^2 - \sqrt{z_m^4-c},~~ 
a^2 = z_m^2 + \sqrt{z_m^4-c}, \\
c &=& [(s^2-M_z^2)^2 - (E+\kappa M_z^2)^2]/(1-\kappa^2), \\
z_m^2 &=& M_z^2 + \frac{s^2-\kappa E}{1-\kappa^2},~~
\tau = \frac{4}{\rho}{\rm K}\left(\frac{z_0}{a}\right),~~
\rho = \sqrt{1-\kappa^2}a, 
\end{eqnarray*}
where sn$(p,x)$, cn$(p,x)$, dn$(p,x)$ are Jacobian elliptic functions and K$(p)$
is a complete elliptic integral.\cite{SO87}

For the case $\kappa=1$ with higher rotational symmetry, considerable
simplifications occur in the classical time evolution.  Both spins precess
uniformly about the direction of the conserved vector ${\bf S}_T \equiv {\bf
  S}_1 + {\bf S}_2$, and the precession rate is $\omega = |{\bf S}_T|$ for both
spins. Equations (\ref{xxzcla}) for the classical actions become
\begin{subequations}\label{hbcla}
\begin{eqnarray}
J_1 &=& 2M_z,\label{hbcla1} \\ 
J_2 &=& \frac{4}{\pi}\int_0^{\pi/2a}\!\!dt\!\!\left[z^2 -
  \frac{z^2s^2+M_z^2-z^2}{(1+\zeta^2)(E+M_z^2-z^2)}\right]\!,\label{hbcla2}
\end{eqnarray}
\end{subequations}
\begin{eqnarray*}
z(t) &=& z_0\sin at,~~ \zeta(t) = \frac{az_0\cos at}{E+M_z^2-z_0^2\sin^2at}, \\
z_0^2 &=&\frac{1}{2}(s^2+E)\left[1- \frac{4M_z^2}{a^2}\right],~~
a = \sqrt{2(s^2-E)}, 
\end{eqnarray*}
and can be evaluated in closed form:
\begin{subequations}\label{hbclaev}
\begin{eqnarray}
J_1 &=& 2M_z,\label{hbcla1ev} \\
J_2 &=& -\sqrt{2(s^2-E)} + (s-M_z){\rm sgn}(s^2-E-2sM_z) \nonumber  \\
&+& (s+M_z){\rm sgn}(s^2-E+2sM_z).\label{hbcla2ev} 
\end{eqnarray}
\end{subequations}
Inverting relations (\ref{hbclaev}) yields a degree-two polynomial dependence of
$E,M_z$ on $J_1,J_2$:
\begin{subequations}\label{hbclainv}
\begin{eqnarray}
I_C(J_1,J_2) &=& M_z = \frac{1}{2}J_1, \label{hbcla1inv} \\
H_C(J_1,J_2) &=& E = s^2 - \frac{1}{2}l_c^2, \label{hbcla2inv}
\end{eqnarray}
\end{subequations}
where $l_c=J_2-|J_1|$ if $s|J_1|>s^2-E$ and $l_c=2s-J_2$ if $s|J_1|<s^2-E$.

%%%%%%%%%%%%%%%%%%%%%%%%%%%%%%%%%%%%%%%%%%%%%%%
%
\subsection{Quantum actions}\label{sec3B}
%
%%%%%%%%%%%%%%%%%%%%%%%%%%%%%%%%%%%%%%%%%%%%%%%
For the case $\kappa=1$, the exact quantum spectrum follows directly from the higher
rotational symmetry of $\hat{H}$:
\begin{equation}\label{hbquml}
\langle\hat{H}\rangle_Q = \hbar^2\sigma(\sigma+1) - 
\frac{\hbar^2}{2}l(l+1),~
\langle\hat{M}_z\rangle_Q = \frac{\hbar}{2} m,
\end{equation}
where $l=0,1,\ldots,2\sigma$ is the quantum number of the total spin and $m= -l,
-l+1,\ldots,+l$ that of its $z$-component. One set of quantum actions (\ref{sclac})
has eigenvalues\cite{note2}
\begin{equation}\label{aqn}
\langle\hat{J}_i\rangle/\hbar \equiv J_i^Q = -\sigma,-\sigma+1,\ldots,+\sigma,
\end{equation}
which are related to $l,m$ as follows:
\begin{subequations}\label{j12qu}
\begin{eqnarray}
J_1^Q &=& \sigma-l,~~ J_2^Q = \sigma-l-m~~~ (m\leq 0), \\
J_1^Q &=& \sigma-l+m,~~ J_2^Q = \sigma-l~~~ (m\geq 0).
\end{eqnarray}
\end{subequations}
The two quantum invariants expressed as explicit functions of action operators
then read
\begin{subequations}\label{hbqujj}
\begin{eqnarray}\label{hbqujj1}
H_Q(\hat{J}_1,\hat{J}_2) = \hat{H} &=& \frac{1}{2}\hbar^2\sigma(\sigma+1) +
\frac{1}{2}{\rm min}(\hat{J}_1,\hat{J}_2) \nonumber \\
 &\times& [\hbar(2\sigma+1) - {\rm min}(\hat{J}_1,\hat{J}_2)], \\
I_Q(\hat{J}_1,\hat{J}_2) = \hat{M}_z &=& 
\frac{1}{2}(\hat{J}_1-\hat{J}_2),
\label{hbqujj2}
\end{eqnarray}
\end{subequations}
where ${\rm min}(\hat{J}_1,\hat{J}_2)$ selects the action operator with the
smaller eigenvalue.

While the functional dependence in (\ref{hbqujj}) is again described by a
degree-two polynomial, it is different from the functional dependence
(\ref{hbclainv}) found classically. The former cannot be reconciled with the
latter by any canonical transformation, nor does the quantum spectrum converge
uniformly toward the classical spectrum for $\sigma\to\infty$, as we shall see in
Sec.~\ref{sec3C1}.

For the cases $0\leq\kappa<1$ we must calculate the $(2\sigma+1)^2$ eigenvalues of the two
quantum invariants $\hat{H}, \hat{M}_z$ by numerical diagonalization of
$\hat{H}$ in the $4\sigma+1$ invariant subspaces of $\hat{M}_z$. From the numerical
data for $\langle\hat{H}\rangle$, $\langle\hat{M}_z\rangle$, we can infer the correct assignment of
action quantum numbers $\langle\hat{J}_i\rangle/\hbar$ to eigenstates by smoothly connecting the
spectrum in parameter space to the known relations (\ref{hbqujj}) for $\kappa=1$.
The resulting data for $H_Q(\hat{J}_1,\hat{J}_2), I_Q(\hat{J}_1,\hat{J}_2)$ can
then be compared with the (semiclassically quantized) inverse classical
relations (\ref{xxzcla}), $H_C(\hat{J}_1,\hat{J}_2)$, $I_Q(\hat{J}_1,\hat{J}_2)$,
to high precision albeit not analytically as in the case $\kappa=1$. Numerical
results will be presented in Sec.~\ref{sec3C2}.

%%%%%%%%%%%%%%%%%%%%%%%%%%%%%%%%%%%%%%%%%%%%%%
%
\subsection{Quantum corrections to quantized actions}\label{sec3C}
%
%%%%%%%%%%%%%%%%%%%%%%%%%%%%%%%%%%%%%%%%%%%%%%
In some simple applications, the functions $H_Q,I_Q$ are identical to the
functions $H_C,I_C$. Hence there are no such quantum corrections. If we take,
for example, the two-spin model $\hat{H} = -\hat{S}_1^z\hat{S}_2^z$, then both
classical invariants $E,M_z$ depend solely on the canonical momenta, and the
latter are identified to be actions: $p_i=J_i$. Hence we have $E=-J_1J_2,
M_z=\frac{1}{2}(J_1+J_2)$, which, upon semiclassical quantization with
$\langle\hat{J}_i\rangle/\hbar = -\sigma,-\sigma+1,\ldots, +\sigma$, yields the exact quantum eigenvalue
spectrum. This situation is exceptional. For all cases of (\ref{Hxxz}) with
$0\leq\kappa\leq 1$, quantum corrections do exist.
%%%%%%%%%%%%%%%%%%%%%%%%%%%%%%%%%%%%%%%%%%%%%%%
%
\subsubsection{Exact results for $\kappa=1$}\label{sec3C1}
%
%%%%%%%%%%%%%%%%%%%%%%%%%%%%%%%%%%%%%%%%%%%%%%%
For the parameter setting $\kappa=1$, the functions $H_Q(\hat{J}_1,\hat{J}_2)$,
$I_Q(\hat{J}_1,\hat{J}_2)$ as given by expressions (\ref{hbqujj}) are to be
compared to the semiclassical expressions $H_C(\hat{J}_1,\hat{J}_2)$,
$I_C(\hat{J}_1,\hat{J}_2)$ inferred from the classical relations
(\ref{hbclainv}) with quantum actions (\ref{aqn}). It turns out to be more
practical to perform the comparison for the inverse functional relations. 
We substitute $\sigma(\sigma+1)$ for $s^2$ and the exact eigenvalues (\ref{hbquml}) for
$E,M_z$ into the classical expressions (\ref{hbclaev}). The result is a set of
non-integer valued semiclassical action quantum numbers 
\begin{subequations}\label{hbsc}
\begin{eqnarray}
J_1^C &=& m, \\
J_2^C &=& \left\{
\begin{array}{ll}
0 & m=l=0 \\
2\sqrt{\sigma(\sigma+1)} - \sqrt{l(l+1)} & |m|<m_0 \\
|m| - \sqrt{l(l+1)} & |m|>m_0, 
\end{array}
\right.
\end{eqnarray}
\end{subequations}
where $m_0=l(l+1)/2\sqrt{\sigma(\sigma+1)}$. An optimal match with the quantum actions
(\ref{j12qu}) can be achieved if we subject (\ref{hbsc}) to two successive
canonical transformations:
\begin{eqnarray*}\label{hbscp}
J_1{^C}' &=& J_1^C, \\ 
J_2{^C}' &=& \left\{
\begin{array}{ll}
2\sqrt{\sigma(\sigma+1)} - |J_1^C| + J_2^C & J_2^C\leq 0 \\ 
J_2^C & J_2^C>0, 
\end{array}
\right.
\end{eqnarray*}
\begin{eqnarray*}\label{hbscpp}
J_1{^C}'' &=& \left\{
\begin{array}{ll}
J_2{^C}' - 2\sqrt{\sigma(\sigma+1)} + \sigma + \frac{1}{2} & J_1{^C}' \leq 0 \\
\nonumber
J_2{^C}' - 2\sqrt{\sigma(\sigma+1)} + \sigma + J_1{^C}' + \frac{1}{2} & J_1{^C}'
 > 0
\end{array}
\right. \\ 
J_2{^C}'' &=& \left\{
\begin{array}{ll}
J_2{^C}' - 2\sqrt{\sigma(\sigma+1)} + \sigma + \frac{1}{2} - J_1{^C}' & J_1{^C}' 
\leq 0 \\
J_2{^C}' - 2\sqrt{\sigma(\sigma+1)} + \sigma + \frac{1}{2} & J_1{^C}' > 0.
\end{array}
\right.
\end{eqnarray*}
We thus arrive at the expressions
\begin{subequations}\label{hbsct}
\begin{eqnarray}
J_1{^C}'' &=& \left\{
\begin{array}{ll}
\sigma + \frac{1}{2} & m=l=0 \\
\sigma - \sqrt{l(l+1)} +\frac{1}{2} & m \leq 0 \\
\sigma - \sqrt{l(l+1)} + \frac{1}{2} + m & m > 0, 
\end{array}
\right. \\
J_2{^C}'' &=& \left\{
\begin{array}{ll}
\sigma + \frac{1}{2} & m=l=0 \\
\sigma - \sqrt{l(l+1)} - m +\frac{1}{2} & m \leq 0 \\
\sigma - \sqrt{l(l+1)} + \frac{1}{2} & m > 0. 
\end{array}
\right.
\end{eqnarray}
\end{subequations}
The deviations of the non-integer valued $J_1{^C}'',J_2{^C}''$ from the integer
valued $J_1^Q,J_2^Q$ then describe the quantum corrections to the semiclassical
actions. 

Using $\sqrt{l(l+1)}-\frac{1}{2} = l+{\rm O}(l^{-1})$, we see at once that the
genuinely quantum mechanical relations (\ref{j12qu}) and the semiclassical
relations (\ref{hbsct}) are asymptotically equivalent at low
energies (large $l$) for $\sigma\to\infty$. At high energies (small $l$), on the other hand,
the two relations remain distinct no matter how large we choose the value of the
spin quantum number $\sigma$.

\begin{figure}[t!]
\centerline{\hspace{1mm}\epsfig{file=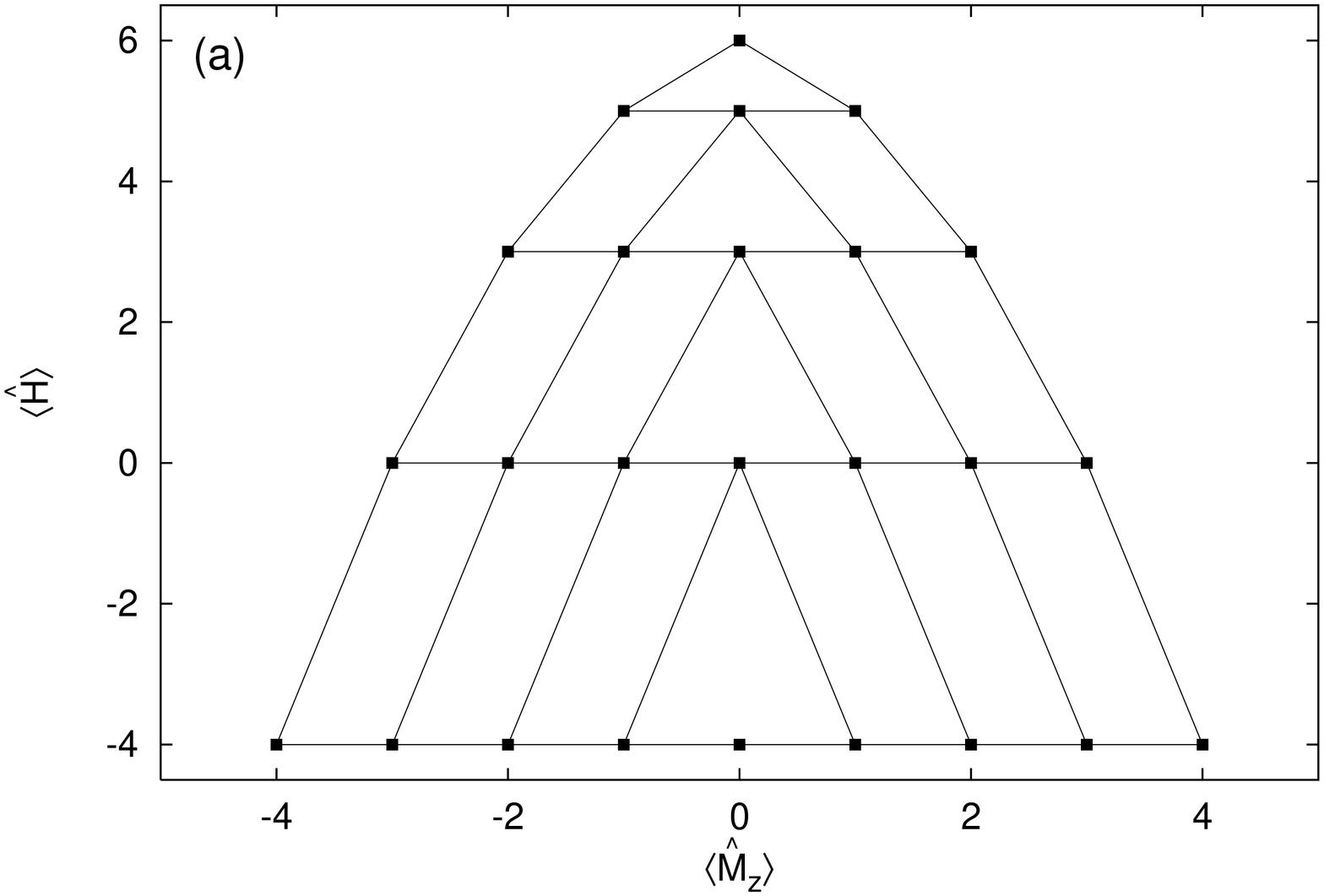,width=5.7cm,angle=-90}}
\centerline{\hspace{1mm}\epsfig{file=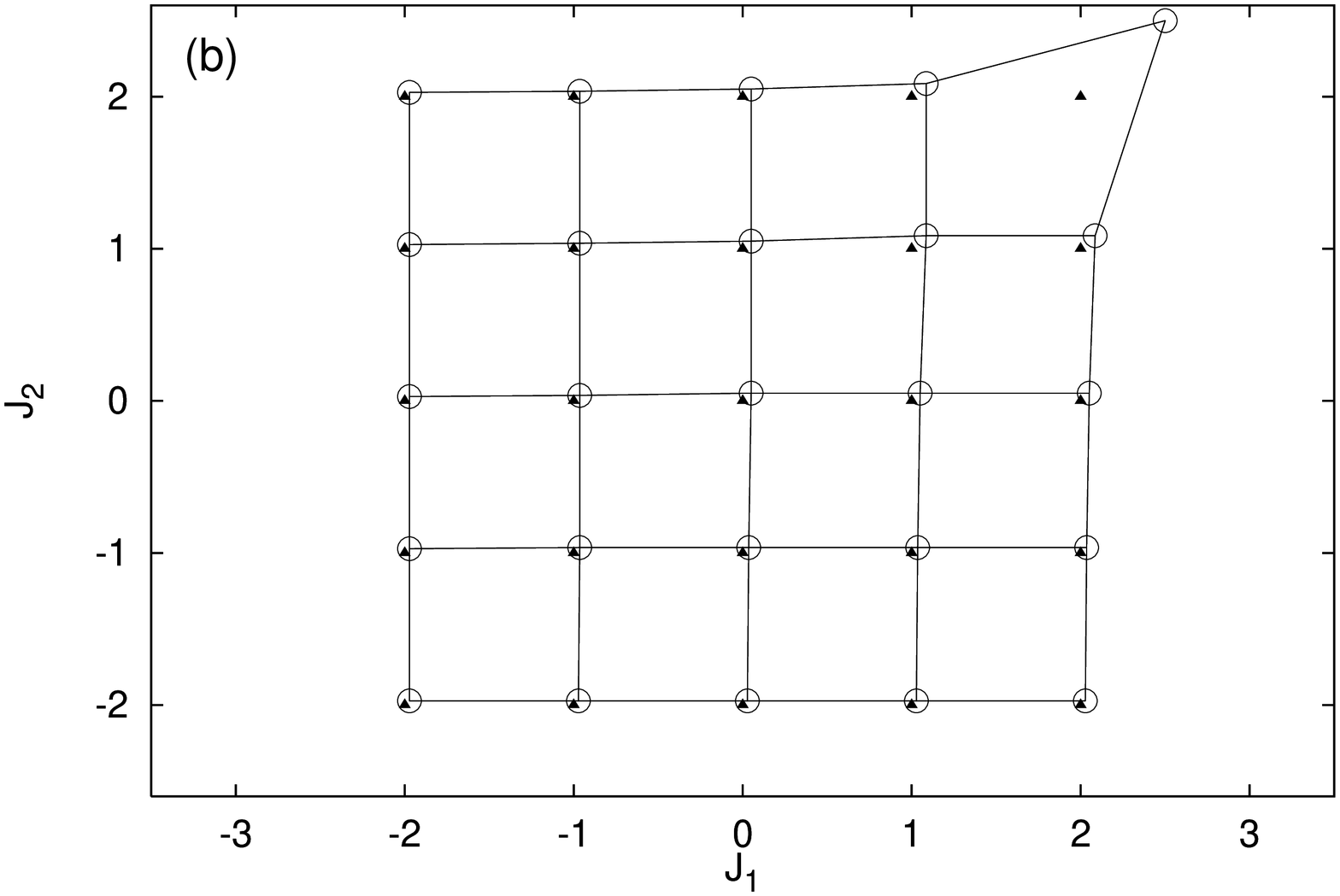,width=5.7cm,angle=-90}}
\caption{(a) Eigenvalue $\langle\hat{H}\rangle$ (energy) versus eigenvalue
  $\langle\hat{M}_z\rangle$ (magnetization) as given in Eqs.~(\ref{hbquml}) of
 all eigenstates of Hamiltonian (\ref{Hxxz}) with
  $\kappa=1$, $\sigma=2$. (b) The full triangles are the quantum images
  $(J_1^Q,J_2^Q)$ of these eigenstates in the action plane as provided by Eqs.
  (\ref{j12qu}). The open circles are the semiclassical images
  $(J_1{^C}'',J_2{^C}'')$ as provided by Eqs.
  (\ref{hbsct}) with $s^2=\sigma(\sigma+1)$. }
\label{fig1}
\end{figure}

\begin{figure}[t!]
\centerline{\hspace{1mm}\epsfig{file=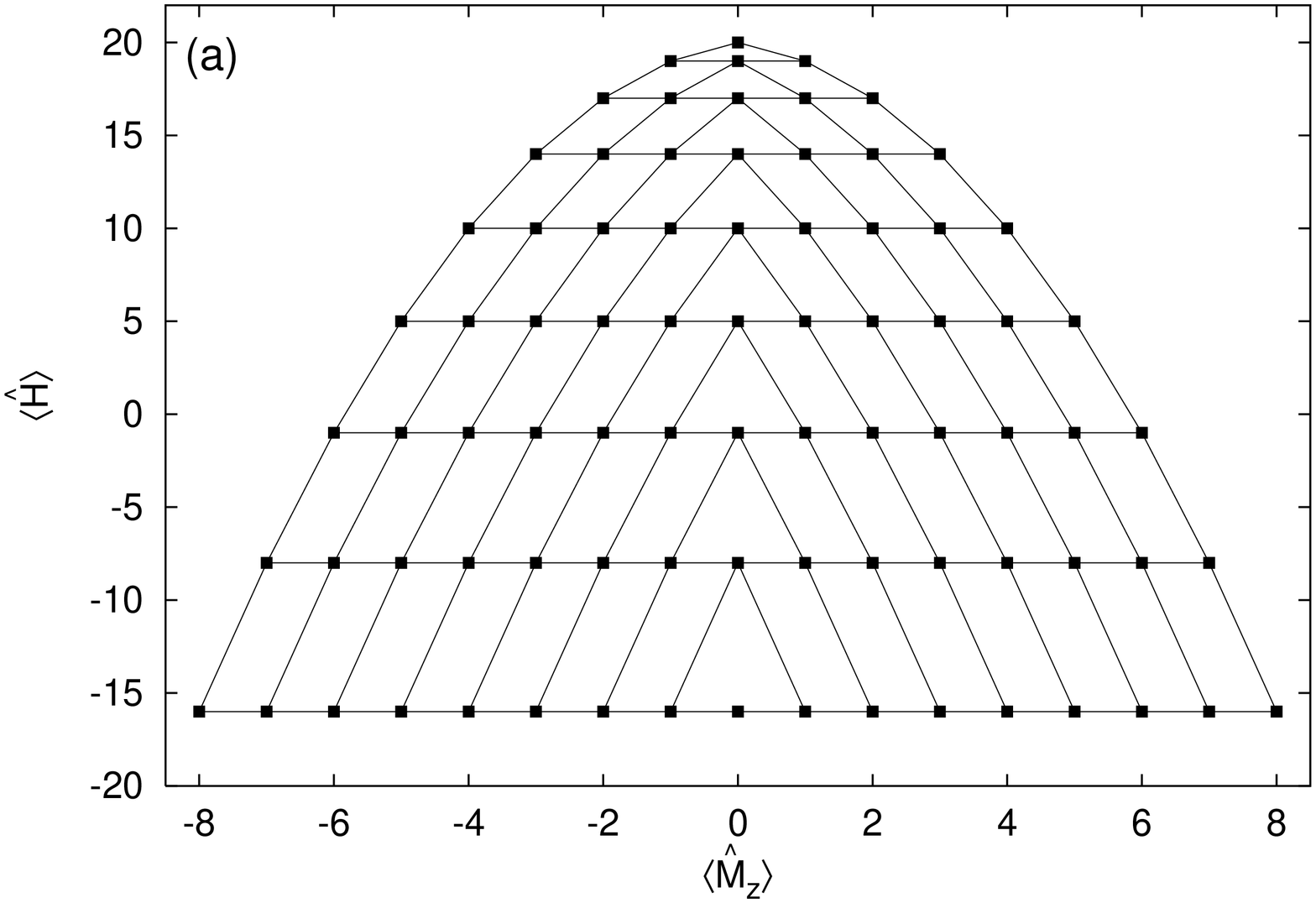,width=5.7cm,angle=-90}}
\centerline{\hspace{1mm}\epsfig{file=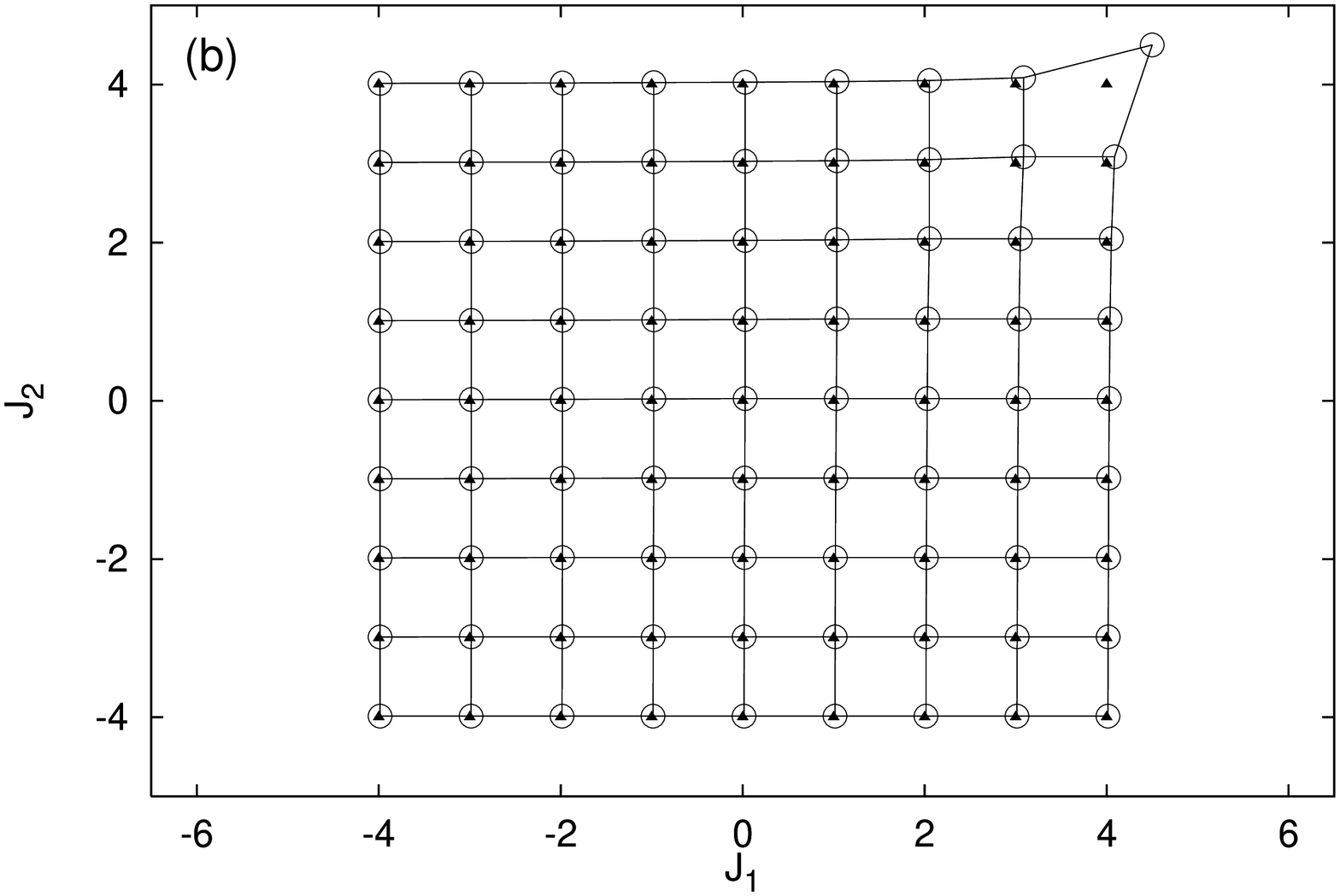,width=5.7cm,angle=-90}}
\caption{Plot of the same quantities as in Fig.~\ref{fig1} but for spin quantum
  number $\sigma=4$.}
\label{fig2}
\end{figure}

To set the stage for the cases $0<\kappa<1$, we plot in Figs.~\ref{fig1}(a) and
\ref{fig2}(a) the eigenvalues of $\hat{H}$ versus those of $\hat{M}_z$ in
representations with spin quantum numbers $\sigma=2$ and $\sigma=4$, respectively. The
patterns of regularity and similarity in the arrays of points are a direct
consequence of the smooth functional relations $H_Q(\hat{J}_1,\hat{J}_2)$,
$I_Q(\hat{J}_1,\hat{J}_2)$. The map $(\langle\hat{H}\rangle, \langle\hat{M}_z\rangle) \to (J_1^Q, J_2^Q)$
from the plane of invariants to the action plane is provided by Eqs.
(\ref{j12qu}) and produces the triangles in Figs.~\ref{fig1}(b) and
\ref{fig2}(b). These points form a perfect lattice with unit spacing.

%%%%%%%%%%%%%%%%%%%%%%%%%%%%%%%BEGIN-FIGURE%%%%%%%%%%
%%%%%%%%%%%%%%%%%%%%%%%%%%%%%%%%END-FIGURE%%%%%%%%K

%%%%%%%%%%%%%%%%%%%%%%%%%%%%%%%BEGIN-FIGURE%%%%%%%%
%%%%%%%%%%%%%%%%%%%%%%%%%%%%%%%%END-FIGURE%%%%%%%

If we use instead the map (\ref{hbsct}) provided by semiclassical quantization,
we obtain the array of open circles in Fig.~\ref{fig1}(b) and
Fig.~\ref{fig2}(b). The bonds shown in parts (a) and (b) of the two graphs
correspond to each other. The distortion in the lattice of circles relative to
the perfect lattice of triangles is a graphical representation of the quantum
corrections in the functions $H_Q(\hat{J}_1,\hat{J}_2)$,
$I_Q(\hat{J}_1,\hat{J}_2)$ relative to the semiclassical functions
$H_C(\hat{J}_1,\hat{J}_2)$, $I_C(\hat{J}_1,\hat{J}_2)$. It visually confirms
what we have already concluded from comparing (\ref{j12qu}) and (\ref{hbsct}),
namely that the deviations die out at low energies (lower left area) but
persist at high energies (upper right area) for $\sigma\to\infty$.
A useful measure of the leading quantum correction to the semiclassical
relation $H_C(\hat{J}_1,\hat{J}_2)$ is the quantity $\sigma\Delta J$, where 
\begin{equation}\label{dsig}
\Delta J \equiv \sqrt{(\Delta J_1)^2 +
  (\Delta J_2)^2},\quad \Delta J_i \equiv J_i^Q -J_i{^C}''
\end{equation}
represents the distance between the triangles and circles on corresponding array
sites in Figs.  \ref{fig1}(b) and \ref{fig2}(b). From Eqs. (\ref{j12qu}) and
(\ref{hbsct}) we obtain
\begin{eqnarray}\label{dj1}
\Delta J = \left\{
\begin{array}{ll}
1/ \sqrt{2} & l=0 \\
\sqrt{2}\left(l-\frac{1}{2}-\sqrt{l(l+1)}\right) & l\neq0 
\end{array} 
\right.
\end{eqnarray}

The dependence of $\sigma\Delta J$ on $J_1^Q, J_2^Q$ thus represents the $1/\sigma$ quantum
correction to the semiclassically quantized actions. It has an inverse first
power divergence in one corner of the action plane for energy levels at the
upper threshold of the spectrum: $\sigma\Delta J \sim [4\sqrt{2}(l/ \sigma)]^{-1}$. For states
with $l/ \sigma\ll 1$ the leading quantum correction is of O(1). In this part of the
spectrum, semiclassical quantization remains inadequate no matter how large we
choose the spin quantum number $\sigma$.

The state with the largest quantum correction to semiclassical quantization is
the singlet combination of the two spins. This state or any nearby state in the
action plane have no proper semiclassical representation.

%%%%%%%%%%%%%%%%%%%%%%%%%%%%%%%%%%%%%%%%%%%%%%%
%
\subsubsection{Numerical results for $0<\kappa<1$}\label{sec3C2}
%
%%%%%%%%%%%%%%%%%%%%%%%%%%%%%%%%%%%%%%%%%%%%%%%%

%%%%%%%%%%%%%%%%%%%%%%%%%%%%%%%BEGIN-FIGURE%%%%%%%%
\begin{figure}[t!]
\centerline{\hspace{1mm}\epsfig{file=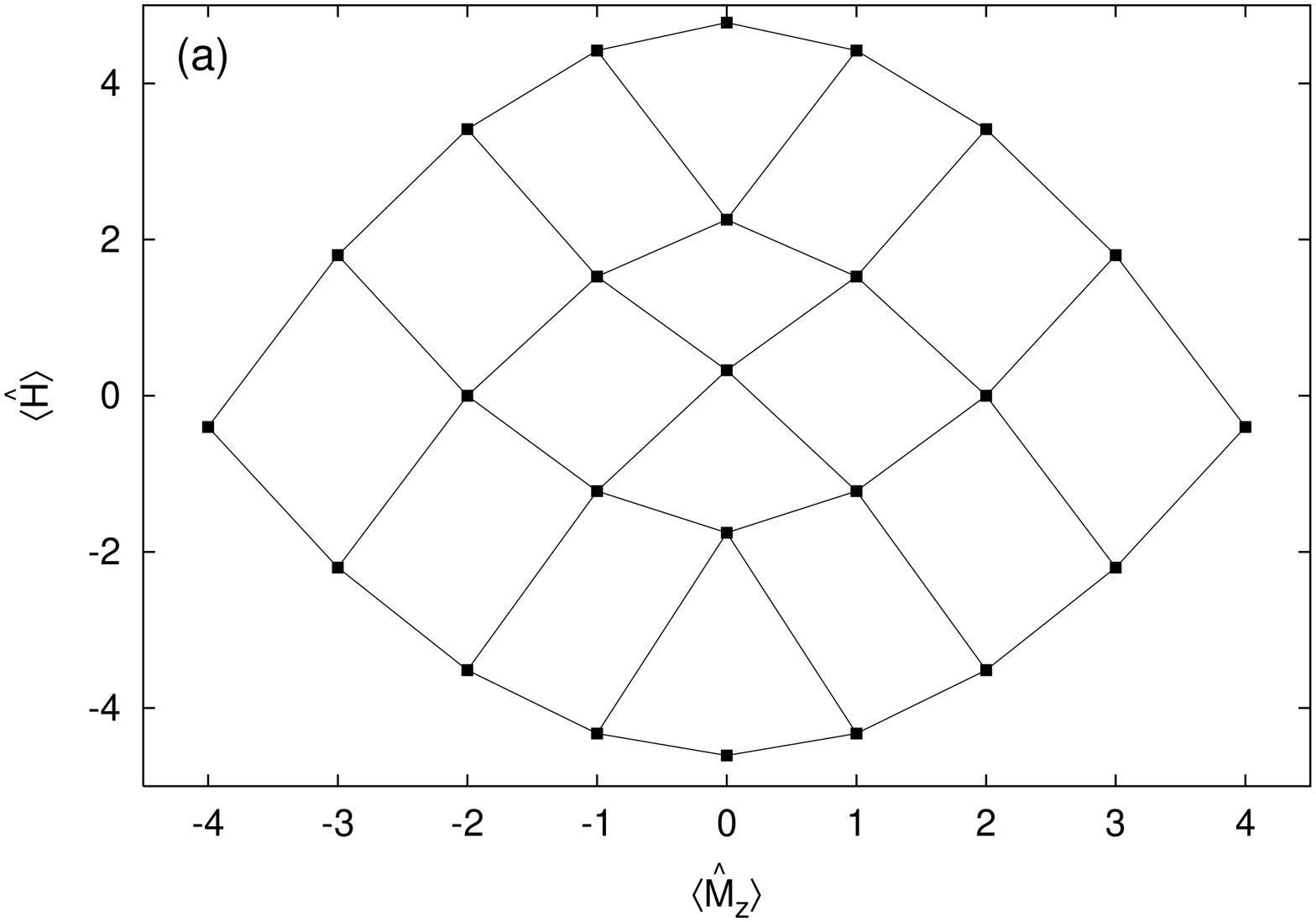,width=5.7cm,angle=-90}}
\centerline{\hspace{1mm}\epsfig{file=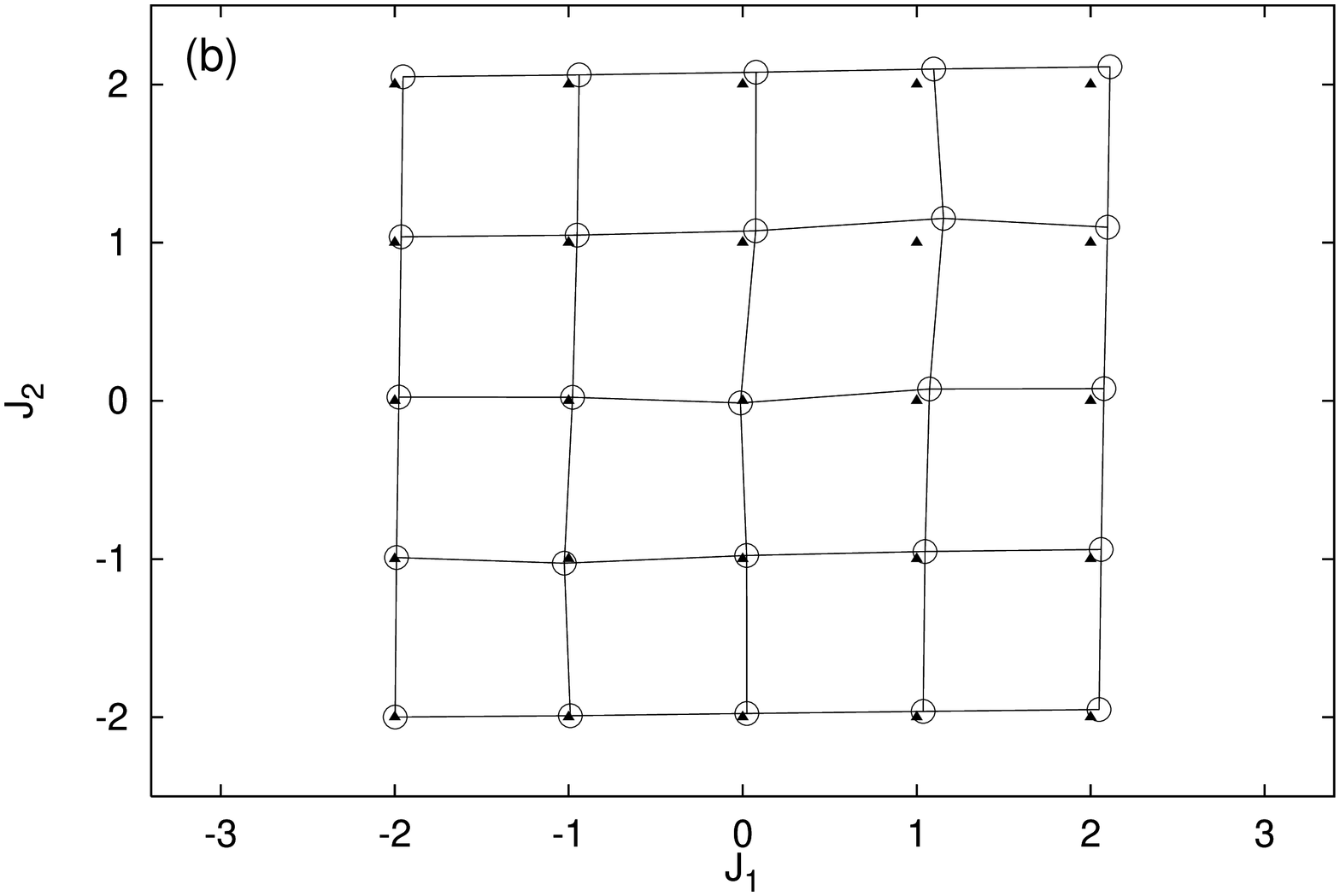,width=5.7cm,angle=-90}}
\caption{(a) Eigenvalue $\langle\hat{H}\rangle$ (energy) versus eigenvalue
  $\langle\hat{M}_z\rangle$ (magnetization) of the $(2\sigma+1)^2=25$ eigenstates of the two-spin
  model (\ref{Hxxz}) with $\kappa=0.1$ for $\sigma=2$. Data from a numerical
  diagonalization. (b) The full triangles are the eigenvalues $J_i^Q=
  \langle\hat{J}_i\rangle/\hbar$ of the action operators, the images of the inverted functions
  $H_Q(\hat{J}_1,\hat{J}_2)$, $I_Q(\hat{J}_1,\hat{J}_2)$. The open circles are
  the semiclassical images $(J_1{^C}'',J_2{^C}'')$ from Eqs.  (\ref{xxzcla})
  with $s^2=\sigma(\sigma+1)$, the images of the inverted functions
  $H_C(\hat{J}_1,\hat{J}_2)$, $I_C(\hat{J}_1,\hat{J}_2)$.}
\label{fig4}
\end{figure}
%%%%%%%%%%%%%%%%%%%%%%%%%%%%%%%%END-FIGURE%%%%%%%%%

Here we use the same graphical representation even though we must rely on the
results of a numerical diagonalization for the energy eigenvalues.  At $\kappa<1$ we
observe that certain features of the quantum invariants change qualitatively
because the rotational symmetry of $\hat{H}$ has been reduced, whereas other
features remain qualitatively the same because the integrability of the model
has not been destroyed.

%%%%%%%%%%%%%%%%%%%%%%%%%%%%%%%BEGIN-FIGURE%%%%%%%%%
\begin{figure}[t!]
\centerline{\hspace{1mm}\epsfig{file=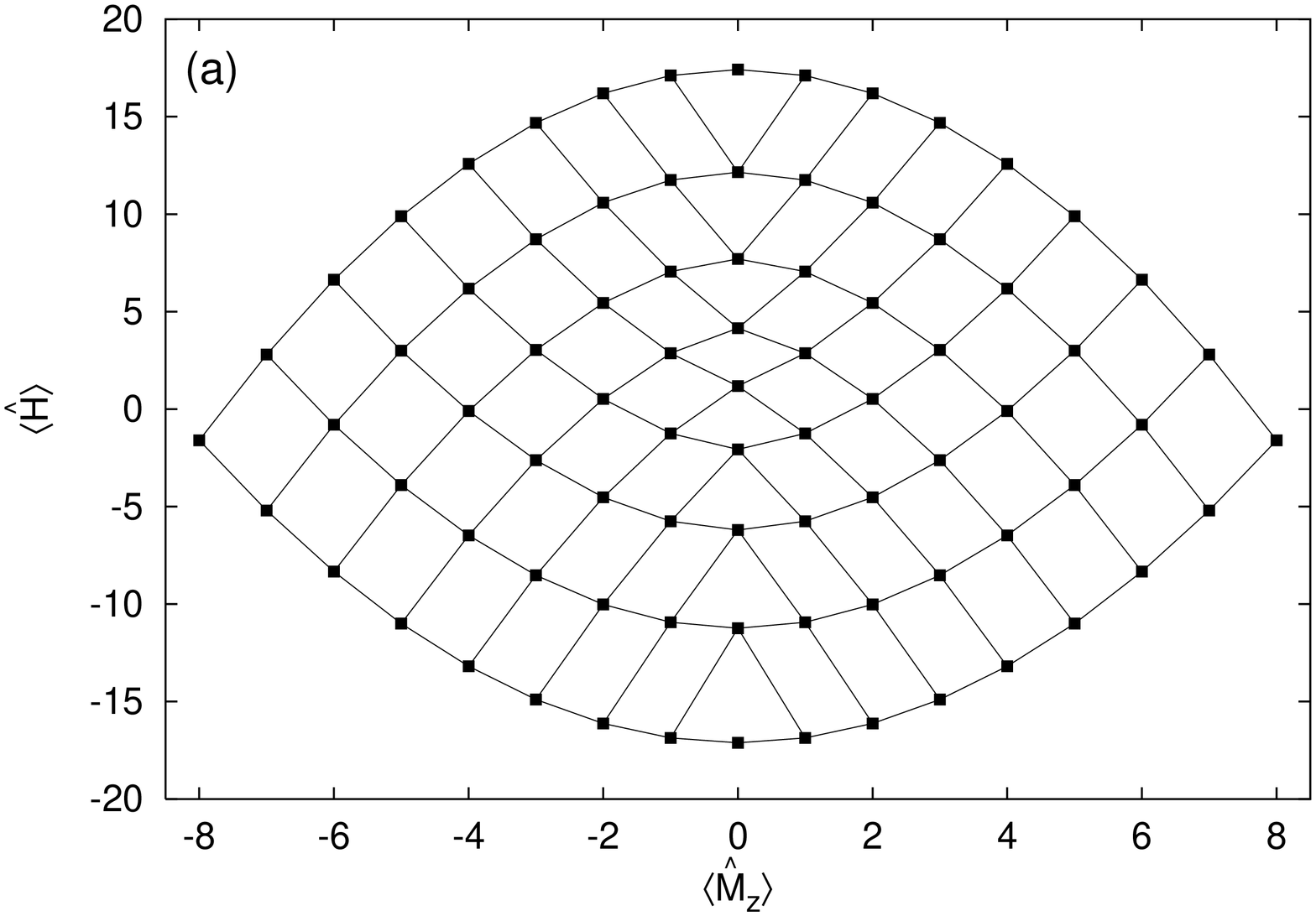,width=5.7cm,angle=-90}}
\centerline{\hspace{1mm}\epsfig{file=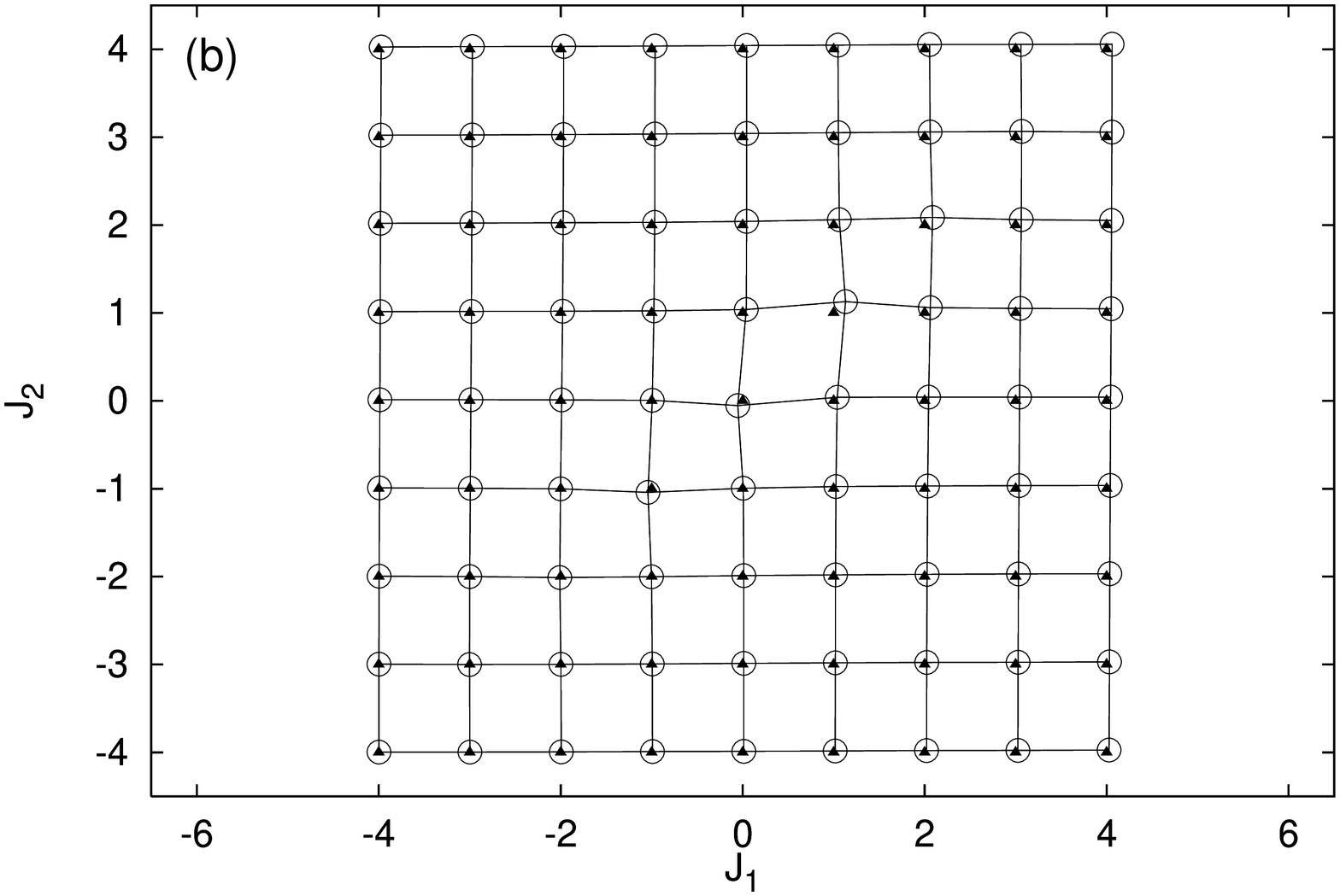,width=5.7cm,angle=-90}}
\caption{Plot of the same quantities as in Fig.~\ref{fig4} but for spin quantum
  number $\sigma=4$. }
\label{fig5}
\end{figure}
%%%%%%%%%%%%%%%%%%%%%%%%%%%%%%%%END-FIGURE%%%%%%%%%

In Figs. \ref{fig4}(a) and \ref{fig5}(a) we have plotted the eigenvalues
$\langle\hat{H}\rangle$, $\langle\hat{M}_z\rangle$ of the two quantum invariants versus each other at
$\kappa=0.1$ for $\sigma=2$ and $\sigma=4$, respectively. Again the data points display regular
patterns. They evolve from the patterns shown in Figs. \ref{fig1}(a) and
\ref{fig2}(a) by smooth deformation of the lines of bonds as the value of $\kappa$ is
lowered gradually. The lower symmetry removes the level degeneracies pertaining
to the strings of horizontal bonds in Figs. \ref{fig1}(a) and \ref{fig2}(a).

When we substitute the eigenvalues $\langle\hat{H}\rangle$ and $\langle\hat{M}_z\rangle$ from the
numerical diagonalization into the exact expression (\ref{xxzcla}) for the
classical actions and subject the resulting set of discrete values $J_i^C$ to
the transformations $J_i^C\to J_i{^C}'\to J_i{^C}''$, we obtain arrays of points in
the form of distorted lattices as illustrated by the open circles in Figs.
\ref{fig4}(b) and \ref{fig5}(b) for the two examples at hand. The deviations of
these data points from the sites of a perfect lattice (marked by triangles)
then again represent the quantum corrections to the (semiclassically) quantized
actions. The patterns in Figs. \ref{fig4}(b) and \ref{fig5}(b) are also
connected to those in Figs. \ref{fig1}(b) and \ref{fig2}(b) by smooth
deformation of the lines of bonds upon gradual variation of the parameter $\kappa$.

\begin{figure}[t!]\vspace*{-1.2cm}
\centerline{\epsfig{file=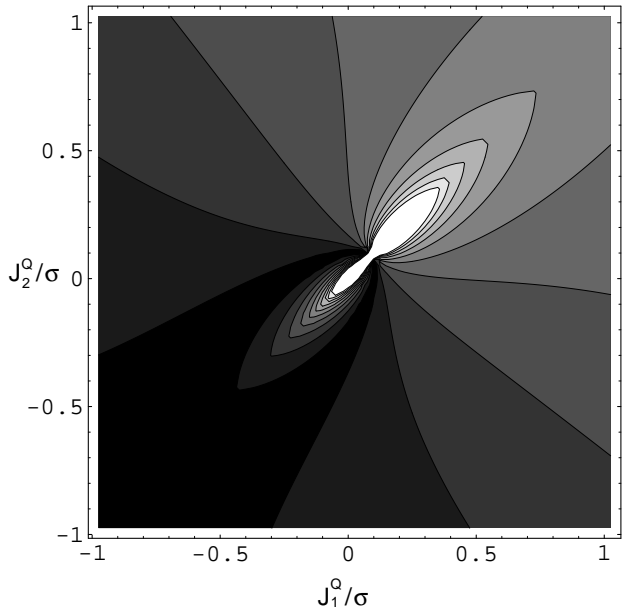,width=7.5cm,angle=0}}
\vspace*{-0.8cm}
\caption{Scaled distance $\sigma\Delta J$ for $\sigma=40$, $\kappa=0.1$ between the images of the
  inverted functions $H_Q(\hat{J}_1,\hat{J}_2)$, $I_Q(\hat{J}_1,\hat{J}_2)$ and
  the images of the inverted functions $H_C(\hat{J}_1,\hat{J}_2)$,
  $I_C(\hat{J}_1,\hat{J}_2)$.}
\label{fig6}
\end{figure}

A closer look at the $1/\sigma$ quantum correction is afforded if we plot the scaled
distance $\sigma\Delta J$ versus the scaled action quantum numbers $J_1^Q/ \sigma$ and $J_2^Q/
\sigma$ for a system with many more levels $(\sigma=40)$. A contour plot of the resulting
landscape is shown in Fig. \ref{fig6}.  Convergence of $\sigma\Delta J$ toward a smooth
function of $J_1^Q/ \sigma, J_2^Q/ \sigma$ is almost uniform. In the case $\kappa=0.1$
considered here, there are two points (as opposed to a single corner point at
$\kappa=1$), where the $1/\sigma$ correction diverges.  The data points $\sigma\Delta J$ closest to
these locations again tend to grow $\propto\sigma$.
%%%%%%%%%%%%%%%%%%%%%%%%%%%%%%%BEGIN-FIGURE%%%%%%%%%
%%%%%%%%%%%%%%%%%%%%%%%%%%%%%%%%END-FIGURE%%%%%%%%

The two sharply peaked maxima in the landscape of Fig.  \ref{fig6} will merge
into a single divergence as $\sigma\to\infty$.  At this point in the action plane, the
leading quantum correction to semiclassical quantization is again of O(1).  Its
location in the action plane does, however, no longer coincide with an extremum
in the energy level spectrum. The divergence in $\sigma\Delta J$ occurs at energy $E=\kappa
s^2$ (for $\sigma\to\infty$), where the classical equations of motion have a fixed point.
For eigenstates with action quantum numbers in the vicinity of this point,
quantum effects persist no matter how large $\sigma$ is made.

\begin{figure}[t!]\vspace*{-1.2cm}
\centerline{\hspace{1mm}\epsfig{file=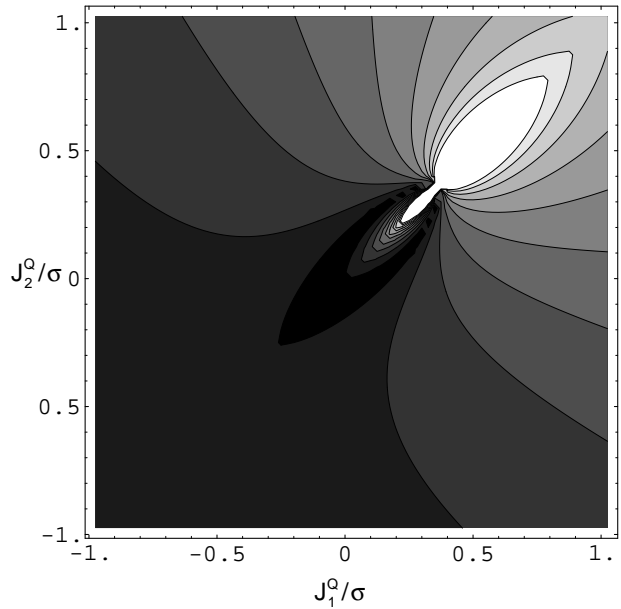,width=7.5cm,angle=0}}
\vspace*{-0.8cm}
\caption{Scaled distance $\sigma\Delta J$ for $\sigma=40$,
  $\kappa=0.5$ between the images of the inverted functions
  $H_Q(\hat{J}_1,\hat{J}_2)$, $I_Q(\hat{J}_1,\hat{J}_2)$ and the images of the
  inverted functions $H_C(\hat{J}_1,\hat{J}_2)$, $I_C(\hat{J}_1,\hat{J}_2)$.}
\label{fig7}
\end{figure}

One point in the action plane where $\sigma\Delta J$ diverges exists throughout the regime
$0\leq\kappa<1$. With $\kappa$ increasing from zero, the singularity moves gradually toward
one corner of the action plane, and the energy of the state pertaining to those
action coordinates moves toward the upper threshold of the spectrum. This trend
is indicated in Fig. \ref{fig7}, which shows the $1/\sigma$-landscape for $\kappa=0.5$.
The endpoint of this gradual shift, the case $\kappa=1$, was described in
Sec.~\ref{sec3C1}.

%%%%%%%%%%%%%%%%%%%%%%%%%%%%%%%BEGIN-FIGURE%%%%%%%%%
%%%%%%%%%%%%%%%%%%%%%%%%%%%%%%%%END-FIGURE%%%%%%%%%

The asymptotic landscape for $\sigma\to\infty$ to which the graphs in Figs.~\ref{fig6} and
\ref{fig7} converge almost everywhere can now be used as the reference frame for
the higher-order quantum corrections. The deviations of the data points from
this new reference, appropriately scaled, will produce another landscape,
representing the $1/\sigma^2$ correction to the semiclassically quantized
actions.\cite{note3} 

\begin{figure}[t!]
\centerline{\hspace{1mm}\epsfig{file=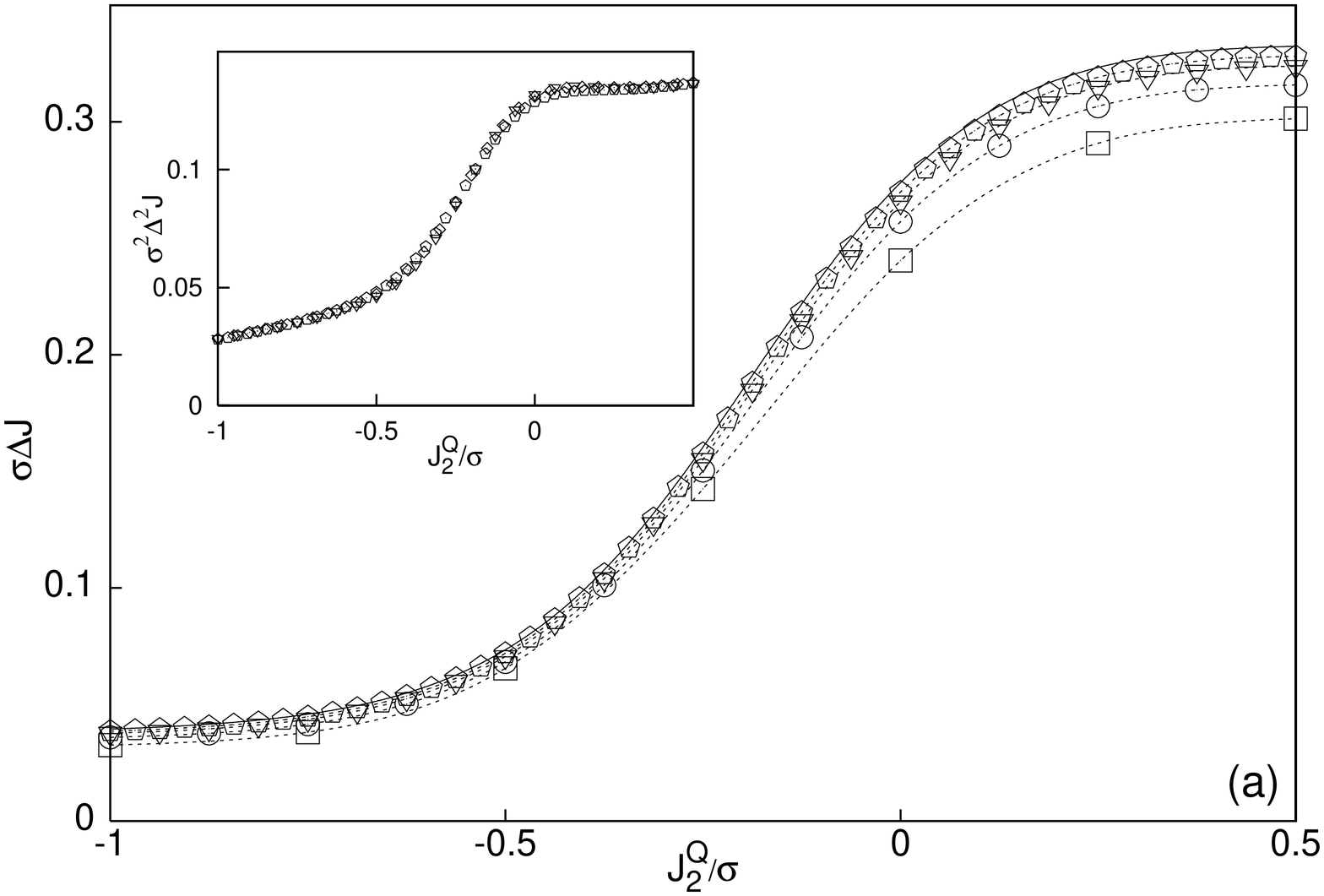,width=5.7cm,angle=-90}}
\centerline{\hspace{1mm}\epsfig{file=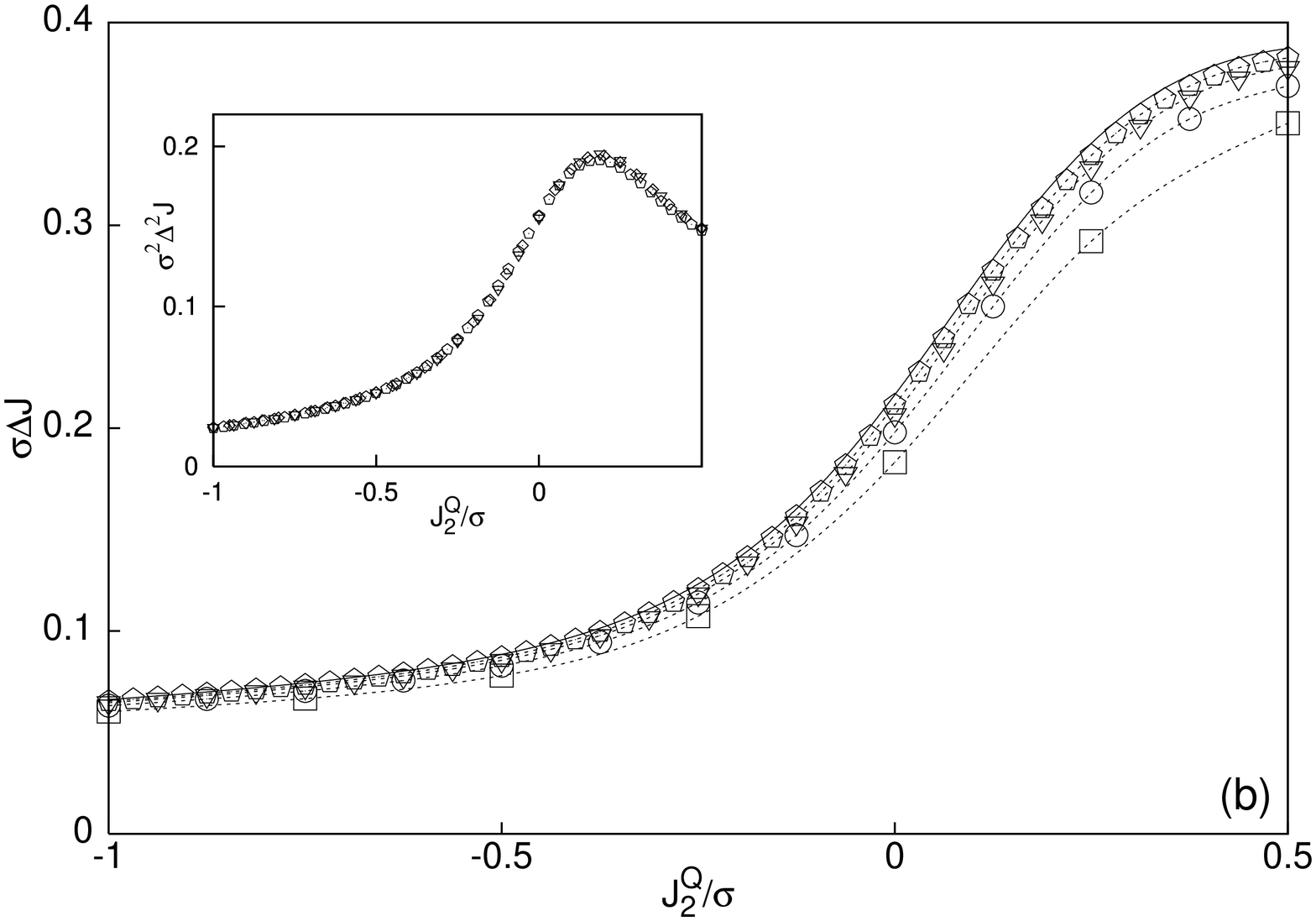,width=5.7cm,angle=-90}}
\vspace*{0.5cm}
\caption{Dependence of the scaled distance $\sigma\Delta J$ for (a) $\kappa=0.1$, (b)
  $\kappa=0.5$ between the images of the inverted functions
  $H_Q(\hat{J}_1,\hat{J}_2)$, $I_Q(\hat{J}_1,\hat{J}_2)$ and the images of the
  inverted functions $H_C(\hat{J}_1,\hat{J}_2)$, $I_C(\hat{J}_1,\hat{J}_2)$.
  Shown are data for $\sigma=4$ (squares), $\sigma=8$ (circles), $\sigma=16$ (triangles),
  $\sigma=32$ (pentagons), and $\sigma=1600$ (solid line). Inset: Scaled deviation $\sigma[\sigma \Delta
  J^{ref} - \sigma \Delta J^{\sigma}]$ of the $\sigma=4,6,8,16$ data from the reference line
  ($\sigma=1600$ data). }
\label{fig8}
\end{figure}

We consider the line $J_2^Q=J_1^Q-\sigma/2$ for this purpose. In the main plot of
Fig.~\ref{fig8} we show the $1/ \sigma$ corrections $\sigma\Delta J$ along this line for
$\sigma=4,8,16,32$. Also shown are data for $\sigma=1600$, which are very close to the
asymptotic values for the $1/ \sigma$ correction and now serve as the
reference line for the $1/ \sigma^2$ corrections.

In the inset to Fig.~\ref{fig8} we have plotted the scaled deviations of the
$\sigma=4,8,16,32$ data from the new reference line. The results suggest that these
data again converge toward a line, which will then be the reference line for $1/
\sigma^3$ corrections. Like the reference line in the main plot of panel (a) [panel
(b)], which is embedded in the landscape Fig.~\ref{fig6} [Fig.~\ref{fig7}], the
new reference line will be embedded in a landscape representing the $1/ \sigma^2$
quantum corrections to semiclassical quantization over the entire action plane.

The point to be emphasized here is not the exact shape of the landscapes that
represent successive orders of quantum corrections to the semiclassically
quantized actions, but that such corrections exist and that the
leading term may be of O(1) at special points rather than of O($\sigma^{-1}$) as
might be expected. 

%%%%%%%%%%%%%%%%%%%%%%%%%%%%%%%BEGIN-FIGURE%%%%%%%%%
%%%%%%%%%%%%%%%%%%%%%%%%%%%%%%%%END-FIGURE%%%%%%%%%

%%%%%%%%%%%%%%%%%%%%%%%%%%%%%%%%%%%%%%%%%%%%%%%%
%
\section{Circular Billiard}\label{sec4}
%
%%%%%%%%%%%%%%%%%%%%%%%%%%%%%%%%%%%%%%%%%%%%%%%%
In the second application we consider a particle of mass $m$ that is
free to move two-dimensionally across a circular area of radius $R$. The
classical Hamiltonian expressed in polar canonical coordinates reads
\begin{equation}
H(p_r,r;p_\vartheta,\vartheta) = \frac{p_r^2}{2m} 
+ \frac{p_\vartheta^2}{2mr^2} + V(r),
\end{equation}
where $V(r)$ is a hard-wall potential that confines the particle to $r\leq R$.

In a recent study, Ree and Reichl\cite{RR98} analyzed this system 
classically and quantum mechanically as an integrable limiting case of the
circular billiard with a straight cut. In general, the cut renders the classical
time evolution chaotic. Here we use some results of Ref.~\onlinecite{RR98} to
investigate the functional dependence of the circular billiard Hamiltonian on
the actions quantum mechanically and semiclassically for comparison with the
two-spin results presented previously.

Integrability of the circular billiard model is guaranteed by the conservation
of angular momentum $L=p_\vartheta$. The canonical transformation to action-angle
coordinates produces the following relations between the integrals of the motion
$E,L$ and the two-action variables:
\begin{subequations}\label{cbj}
\begin{eqnarray}\label{cbj1}
J_1 &=& L, \\ \label{cbj2}
J_2 &=& \frac{\sqrt{2mE}}{\pi}\left[\sqrt{R^2-x^2} - 
x\arccos\left(\frac{x}{R}\right)\right],
\end{eqnarray}
\end{subequations}
where $x=\sqrt{L^2/2mE}$. The eigenfunctions of the circular billiard, i.e. the
solutions of 
\begin{equation}
\left(\frac{\partial^2}{\partial r^2} + \frac{1}{r}\frac{\partial}{\partial r}
+ \frac{1}{r^2}\frac{\partial^2}{\partial\vartheta^2} + k^2\right)
\Psi(r,\vartheta) = 0
\end{equation}
with $k^2=2mE/\hbar^2$ and Dirichlet boundary conditions are known. The exact
expressions for the two quantum invariants $\hat{H}$ (energy) and $\hat{L}$
(angular momentum) are
\begin{equation}\label{cbel}
\langle\hat{H}\rangle = \frac{\hbar^2\alpha_{lk}^2}{2mR^2},~~~ \langle\hat{L}\rangle = \pm l\hbar,
\end{equation}
where $l=0,1,2,\ldots$ and $\alpha_{lk}$ is the $k^{\rm th}$ zero $(k=1,2,\ldots)$ of the Bessel
function J$_l(x)$.

One major distinction between the circular billiard model and the two-spin model
is that all invariant Hilbert subspaces are infinite-dimensional in the former
and finite-dimensional in the latter. The energy has no upper bound in the
circular billiard and the angular momentum neither upper nor lower bound.

In Fig. 9 we have plotted the eigenvalues $\langle\hat{H}\rangle$ versus $\langle\hat{L}\rangle$ of the
two quantum invariants near the bottom of the level spectrum. As in the two-spin
model, the regular pattern of points is a signature of quantum integrability. In
both models the points tend to become displaced irregularly when nonintegrable
perturbations are introduced.\cite{SM90,RR98}

%%%%%%%%%%%%%%%%%%%%%%%%%%%%%%%BEGIN-FIGURE%%%%%%%%%
\begin{figure}[t!]\hspace*{-0.5cm}
\centerline{\epsfig{file=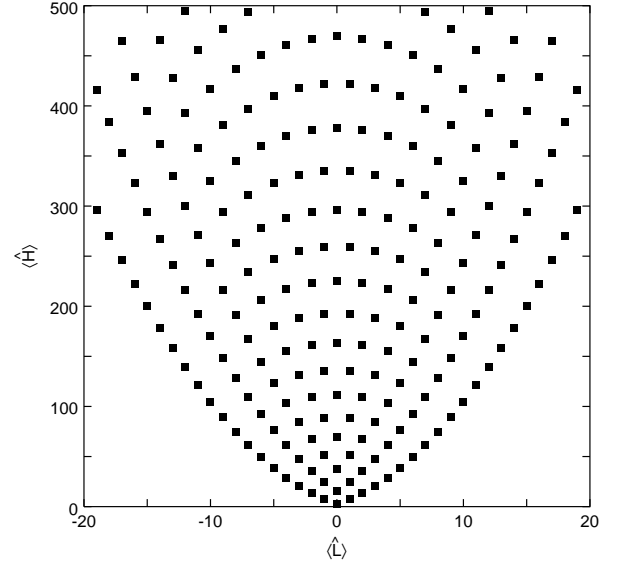,width=8.0cm,angle=-90}}
\caption{Eigenvalue $\langle\hat{H}\rangle$ (energy) versus eigenvalue $\langle\hat{L}\rangle$ (angular
  momentum) as given in Eq.~(\ref{cbel}) of the eigenstates near the bottom of
  the spectrum of the circular billiard model.}
\label{fig9}
\end{figure}
%%%%%%%%%%%%%%%%%%%%%%%%%%%%%%%%END-FIGURE%%%%%%%%%

The integers $k,l$ in (\ref{cbel}) can be identified as the eigenvalues (in
units of $\hbar$) of a set of quantum actions:
\begin{equation}\label{cbqa}
\langle\hat{J}_1\rangle=\hbar l,~ ~ ~ \langle\hat{J}_2\rangle=\hbar(k-\frac{1}{4}).
\end{equation}
The shift in the second expression is dictated by a Maslov index $\alpha_1=1$ (see
Sec.~\ref{sec2}).\cite{Perc77}  The results (\ref{cbel}) combined with (\ref{cbqa}) thus
define specific functional relations $H_Q(\hat{J}_1,\hat{J}_2)$,
$I_Q(\hat{J}_1,\hat{J}_2)$ between quantum invariants and quantum actions. They are
to be compared with the functional relations $H_C(\hat{J}_1,\hat{J}_2)$,
$I_C(\hat{J}_1,\hat{J}_2)$ as defined by (\ref{cbj}) combined with
(\ref{cbqa}).

\begin{figure}[t!] \vspace*{-1.7cm}\hspace*{0.3cm}
\centerline{\hspace{1mm}\epsfig{file=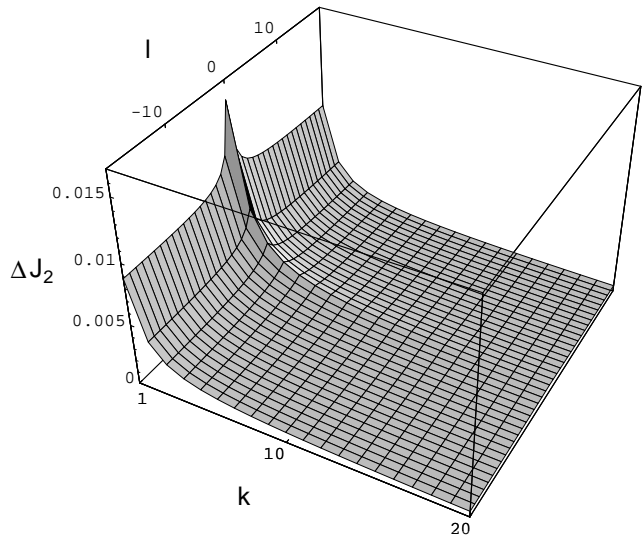,width=8.0cm,angle=0}}
\vspace*{-2.0cm}
\caption{Quantum corrections to the semiclassical prediction for the energy
  eigenvalues of the circular billiard model. Plotted is the deviation $\Delta J_2 =
  |J_2^Q-J_2^C|$, where $J_2^Q=k-\frac{1}{4}$ and $J_2^C=J_2/ \hbar$ as determined
  by (\ref{cbj2}) with $E=\langle \hat{H}\rangle, L=\langle \hat{L}\rangle $ substituted from (\ref{cbel}).}
\label{fig10}
\end{figure}

For a graphical representation of the quantum corrections to semiclassical
quantization, we proceed as in Sec.~\ref{sec3}. In Fig.~\ref{fig10} we plot $\Delta
J_2\equiv|J_2^Q-J_2^C|$ versus $k$ and $l$, where $J_2^Q=k-\frac{1}{4}$ and $J_2^C$
is the value of (\ref{cbj2}) when the exact eigenvalues (\ref{cbel}) for the
quantum invariants are substituted into the expression.

We observe a landscape in the form of a sloped ridge centered at $l=0$. The
largest quantum correction to semiclassical quantization pertains to the ground
state (with $k=1,l=0$). The plot suggests that the quantum
corrections die out for large $k$. This is confirmed by substitution of the 
asymptotic expression for $k \gg l$,\cite{SO87}
\begin{equation}\label{cbask}
\alpha_{lk} \sim \beta - \frac{4l^2}{8\beta} + {\rm O}(\beta^{-3}),~ \beta = k+\frac{l}{2}-\frac{1}{4},
\end{equation}
into (\ref{cbel}) for use in (\ref{cbj2}):
\begin{equation}\label{cbj2ask}
J_2(l,k) \sim \hbar\left[k - \frac{1}{4} + \frac{1}{8\pi^2k} + {\rm O}(k^{-2})\right],~ 
k\gg l.
\end{equation}
The quantum corrections also decrease with increasing $|l|$ at fixed $k$, but
not all the way to zero. To demonstrate this for $k=1$, we use the asymptotic
expression for $l\gg k=1$,\cite{SO87}
\begin{equation}\label{cbasl}
\alpha_{l1} \sim |l| + C_1|l|^{1/3} + C_2|l|^{-1/3}
\end{equation}
with $C_1\simeq 1.8558$, $C_2\simeq 1.033$ for use in (\ref{cbel}). When substituted into
(\ref{cbj2}) we obtain the asymptotic value
\begin{equation}\label{cbj2asl}
J_2(l,1) = (\hbar/3\pi)(2C_1)^{3/2} + {\rm O}(|l|^{-2/3}),
\end{equation}
which deviates from the reference value $\hbar(1-\frac{1}{4})$ by roughly one
percent. The conclusion is that the semiclassical regime of the circular billiard
is restricted to states with $k\gg l$. It does not include, for example, any
states along the lowest branch $(k=1)$ shown in Fig.~\ref{fig10}, no matter how
large the energy of the state becomes with increasing $|l|$.

%%%%%%%%%%%%%%%%%%%%%%%%%%%%%%%BEGIN-FIGURE%%%%%%%%%
%%%%%%%%%%%%%%%%%%%%%%%%%%%%%%%%END-FIGURE%%%%%%%%%

%%%%%%%%%%%%%%%%%%%%%%%%%%%%%%%%%%%%%%%%%%%%%%%%%
%
\section{Conclusion}\label{sec6}
%
%%%%%%%%%%%%%%%%%%%%%%%%%%%%%%%%%%%%%%%%%%%%%%%%%
In this study we have investigated a key signature of quantum integrability in
systems with two degrees of freedom, namely the functional dependence of the
Hamiltonian $\hat{H}$ and the second integral of the motion $\hat{I}$ on two
action operators $\hat{J}_1, \hat{J}_2$. 

The results presented in Secs.~\ref{sec3} and \ref{sec4} for the
(semiclassically) quantized and the (primary) quantum energy level spectra of
two integrable model systems suggest the following interpretation, which is
consistent with the conclusions inferred from an entirely different line of
reasoning:\cite{WM95} (i) Quantum integrability implies that the Hamiltonian can
be expressed as an operator valued function of the actions: $\hat{H} =
H_Q(\hat{J}_1,\hat{J}_2)$, where the eigenvalue spectrum of the action operators
is of the form (\ref{sclac}).  (ii) This function is different from the function
$H_C(\hat{J}_1,\hat{J}_2)$ inferred via semiclassical quantization from the
solution of the classical dynamical problem.  (iii) In some asymptotic regime
associated with the classical limit the function $H_Q(\hat{J}_1,\hat{J}_2)$
converges, if properly scaled, toward the function $H_C(\hat{J}_1,\hat{J}_2)$,
but the convergence need not be uniform. (iv) For the second integral of the
motion, which (classically) guarantees integrability, there exist functions
$I_Q(\hat{J}_1,\hat{J}_2)$ and $I_C(\hat{J}_1,\hat{J}_2)$ with analogous
properties.

The existence of action operators as constituent elements of all quantum
invariants in integrable model systems is a key property necessary to explain
the dimensionality of level crossing manifolds relative to the dimensionality of
integrability manifolds in the parameter space of model systems with parametric
integrability conditions.  On the $d_I$-dimensional integrability manifold in
the parameter space of a given model system, both functions
$H_Q(\hat{J}_1,\hat{J}_2)$ and $I_Q(\hat{J}_1,\hat{J}_2)$ will then depend
continuously on these parameters. The quantum eigenvalue spectrum on the
integrability manifold is determined by $\langle\hat{H}\rangle_Q =
H_Q(\langle\hat{J}_1\rangle,\langle\hat{J}_2\rangle)$ and can be interpreted as a set of continuous
functions of the Hamiltonian parameters subject to the constraints imposed by
the integrability condition. The level crossings, which occur at the
intersections of the graphs of any two members from the set of functions are
then naturally confined to $(d_I-1)$-dimensional manifolds and are naturally
embedded in the integrability manifold, in agreement with empirical
evidence.\cite{SM98}

In a companion paper,\cite{SM99b} we have investigated
how the existence of $H_Q(\hat{J}_1,\hat{J}_2), I_Q(\hat{J}_1,\hat{J}_2)$ within
the five-dimensional integrability manifold of a six-parameter two-spin model
affects the properties of quantum invariants and what impact on the same
quantities the nonexistence of $H_Q(\hat{J}_1,\hat{J}_2)$,
$I_Q(\hat{J}_1,\hat{J}_2)$ elsewhere in parameter space has.

\acknowledgments
This work was supported by the Research Office of the University of Rhode Island.
We are very grateful to Joachim Stolze for his comments and suggestions relating
to this work.

%%%%%%%%%%%%%%%%%%%%%%%%%%%%%%%%%%%%%%%%%%%%%%%%


\begin{thebibliography}{99}
\bibitem{note1} The exchange constant of this model, which is measured in
  arbitrary energy units divided by $\hbar^2$, has been suppressed to avoid a
  cluttered notation.
\bibitem{note2} Equivalent sets of actions can be obtained by sums or
  differences of the actions (\ref{aqn}) with arbitrary additive constants.
\bibitem{note3}  For the case $\kappa=1$, all quantum corrections can be
extracted analytically to all orders from Eqs.~(\ref{dj1}).
\end{thebibliography}
\end{document}